\documentclass[fleqn,runningheads]{llncs}
\usepackage{amsmath}
\usepackage{mathtools}

\begin{document}

\title{Graph? Yes! Which one? Help!}

\author{Ora Lassila \and Michael Schmidt \and Brad Bebee \and Dave Bechberger \and Willem Broekema \and Ankesh Khandelwal \and Kelvin Lawrence \and Ronak Sharda \and Bryan Thompson}

\authorrunning{Lassila et al}

\institute{Amazon Web Services\footnote{Copyright \copyright{} 2021 Amazon Web Services. Use permitted under a CC BY 4.0 license.}, Seattle, WA\\
\email{ora@amazon.com}}

\maketitle

\begin{abstract}
Amazon Neptune is a graph database service that supports two graph (meta)models: W3C’s Resource Description Framework (RDF) \cite{Klyne+Carroll-RDFSpec2004} and Labeled Property Graphs (LPG) \cite{rodriguez2010,robinson2013}. Customers opt in for one or the other model, and this choice determines which data modeling features can be used, and – perhaps more importantly – which query languages are available to query and manipulate the graph. The choice between the two technology stacks is difficult and requires consideration of data modeling aspects, query language features, their adequacy for current and future use cases, as well as many other factors (including developer preferences). Sometimes we see customers make the {\em wrong} choice with no easy way to reverse it later.

It is therefore highly desirable that the choice of the query language can be made without consideration of what graph model is chosen, and can be easily revised or complemented at a later point. In this paper, we advocate and explore the idea of a single, unified graph data model that embraces both RDF and LPGs, and naturally supports different graph query languages on top. We investigate obstacles towards unifying the two graph data models, and propose an initial unifying model, dubbed ``one graph''
(``1G'' for short), as the basis for moving forward.

\end{abstract}

\section{Introduction}

The Amazon Neptune graph database supports both RDF and LPG
models and query languages, but today customers are not able to “cross-use” these technologies (e.g., a user cannot query RDF data using Gremlin). This as well as the differences in domain modeling approaches between the two graph models cause confusion among prospective Neptune users. These users frequently struggle with the choice between the two models. We have even seen customers make the {\em wrong} choice with no easy way to reverse direction later. Occasionally, we see requests for interoperability (i.e., cross-use) from existing users who are already proficient with RDF or LPG, or both, and from customers with use cases that are sometimes better served via RDF and at other times better served via LPG.

Making a good choice between the two technology stacks is complex and requires a balanced consideration of data modeling aspects, query language features, and their adequacy for current and future use cases. Furthermore, the choice has impact on aspects such as ease of data maintenance, augmentation and interchange, software ecosystems (such as explorers, visualization and application development tools), or simply developer expertise and preferences. We have repeatedly seen our customers struggle with this decision, and the eventual choice is often dominated by the query language that appears most appealing for implementing the initial use case, yet additional requirements and new use cases that emerge over time may challenge the initial choice.

RDF offers a formal model that supports global identifiers (IRIs), well-defined graph merging, a natural way to break a graph into subgraphs, and federated query, making it well-suited to use cases where multiple (potentially external) data sources are used (see, for example, the Linked Open Data project\footnote{\url{https://lod-cloud.net/}}). RDF also supports the reuse of schemas in the form of vocabularies for ontology definition as well as logical reasoning. Not surprisingly, we often see information architects prefer the features of the RDF model because of a good fit with use cases for data alignment, master data management, and data exchange. LPGs, on the other hand, are more in line with familiar programming models, and offer good integration with a number of programming languages. Software developers often choose an LPG language because they find it more natural and more “compatible” with their programming paradigm. Developers coming from the SQL world often like that a vertex in an LPG is much like a row in a relational database. Unfortunately, many of these developers are then challenged when they need to expand a single graph to multiple data sources or when they have to deal with external data. Note that the choice of LPG can also happen when RDF is dismissed out of hand because it is viewed as complex and ``academic''. Regardless of what the reasons, we believe that the (forced) choice of graph models slows the adoption of graphs because it creates confusion and segmentation in the graph database space.

What if users could choose between different query languages, independent of what graph model they have decided to use? From interactions with Neptune customers we have noticed that there are often strong preferences for a particular query language, and there are also situations where one query language is simply better suited because of its particular features (e.g., expressiveness of graph traversals and path queries in Gremlin vs. \texttt{SERVICE} federation in SPARQL \cite{Prudhommeaux+Buil-Aranda-SPARQLSpec2013}). We argue that one’s choice of a query language should not dictate that one has to forego benefits of a graph model said query language does not support.

Therefore, in this paper we examine the idea of {\em graph interoperability}. That is, removing the obstacles that prevent us from using SPARQL \cite{Harris+Seaborne-SPARQLSpec2013} over LPGs, Gremlin \cite{rodriguez2015} or openCypher\footnote{\url{https://github.com/opencypher}} over RDF, etc. The goal is not merely to be able to cross-use query languages, but to be able to do it in a manner where the user does not have to be cognizant (and careful) about how the interoperability is achieved. In other words, we are interested in a data model that combines both RDF and LPG into a model providing a unified semantics that includes (and generalizes) the specifics of the individual models. In addition to flexibility in choosing the query language, the idea is that this unified data would also allow graph users to combine and interlink data sets maintained in both RDF and LPG formats. More generally, the idea of providing a unified data model abstracts away the need for customers to choose a data format ahead of time, therefore removing a major obstacle to graph database adoption.

In this context it should be noted that less interesting is some kind of “qualified interoperability” where the cross-use of query languages would require one to understand the implementation of the underlying graph models – in other words, we want to stay away from, say, {\em implementing} RDF using an LPG,\footnote{See, for example, this: \url{https://www.tigergraph.com/blogs/about-tigergraph/graph-gurus-episode-1-building-an-enterprise-knowledge-graph-from-rdf-data/}} because this simply requires the users to be knowledgeable of multiple models instead of a single one.

\section{The 1G Model}
\label{1G}

We start our investigation in the context of a graph database (Neptune) that uses a ``quad-based'' internal representation for both RDF and for LPG, and while this may seem to bias our thinking towards RDF-centric solutions, our broad goal is to not be preferential to either graph model.

In the 1G data model, we represent data -- RDF graphs, LPGs, or both -- using a set of so-called {\em statements} of the form
\begin{align*}
\texttt{src} \xrightarrow{\text{label}} \texttt{value} : sid
\end{align*}
where \texttt{src} represents a source vertex, \texttt{label} denotes the edge or property label, \texttt{value} denotes either the target vertex or the property value (such as a string, number, etc.), and $sid$ assigns a globally unique ID (as will be discussed later, the same statement might come with different identifiers). For RDF, 1G statements map directly to SPO triples with unique ID; for LPGs, they give us a unified mechanism to represent edges and properties alike. If the type of \texttt{value} is fixed (to either a vertex or a property value), we may refer to statements as {\em edges} and {\it properties}, respectively. The main purpose of this formalism is to define the semantics of our model and does not imply a particular physical implementation. 

The $sid$s  can be used in either \texttt{src} or \texttt{value} position, to refer to the statements that they represent (say, to make statements about these statements). As a concrete example for a 1G dataset, consider the following statements:
\begin{align*}
&\texttt{Alice} \xrightarrow{\text{knows}} \texttt{Bob} : sid_1\\
&\texttt{Alice} \xrightarrow{\text{name}}  \texttt{"Alice"} : sid_2\\
&\texttt{Bob}  \xrightarrow{\text{name}} \texttt{"Bob"} : sid_3
\end{align*}
We can then introduce additional statements that refer to the ones above by referencing existing $sids$, in this example $sid_1$, for instance:
\begin{align*}
&sid_1 \xrightarrow{\text{since}} 2020 : sid_4
\end{align*}

Conceptually, we distinguish between {\em ground} statements (or ground edges and properties, if their \texttt{value} type is known) as statements that do not contain $sids$, and {\em assertions}, represented as statements with $sid$s in \texttt{src} and/or \texttt{value} position.\footnote{We do not see a strong use case for supporting $sid$s in the \texttt{label} position.} In the previous example, our three ground statements describe two persons Alice and Bob with their names, stating that Alice knows Bob. The assertion over the $sid$ of the first ground statement states that the \texttt{knows} relationship between Alice and Bob, represented by $sid_1$, exists since 2020 -- this is how 1G captures what is commonly referred to as {\em edge properties} in LPGs.

Interoperability between graphs can now be defined as ``views'' of the 1G model for RDF, RDF-star \cite{hartig2021}, and LPG. Those views can be understood as specific interpretations (or mappings) of the generic 1G model into a specific graph data model. We sketch this idea by providing possible interpretations for our toy graph, for RDF, RDF-star, and LPG, respectively:
\begin{verbatim}
# Plain RDF without reification (no support for edge properties)
:Alice :knows :Bob .
:Alice :name "Alice" .
:Bob :name "Bob" .

# RDF-star: edge properties = statements about "reified" triples
:Alice :knows :Bob .
<<:Alice :knows :Bob>> :since 2020 .
:Alice :name "Alice" .
:Bob :name "Bob" .
\end{verbatim}
With LPG, assertions over edges are viewed as edge properties, so our LPG interpretation of the 1G model consists of the following vertices and edges:
\begin{align*}
&v_1 \textrm{ with } id = \texttt{Alice} \textrm{ and properties } \{ \texttt{name} => \texttt{"Alice"} \}\\
&v_2 \textrm{ with } id = \texttt{Bob} \textrm{ and properties } \{ \texttt{name} => \texttt{"Bob"} \}\\
&e_1 \textrm{ from } v_1 \textrm{ to } v_2 \textrm{ with label } \texttt{knows} \textrm{ and properties } \{ \texttt{since} => 2020 \}
\end{align*}
There exist, of course, other possible interpretations of the 1G model for each of the existing graph data formats (and we deliberately omit details such as mapping between 1G identifiers and RDF/LPG identifiers, which we will discuss in Section~\ref{merging}). For instance, in our example we chose the assertion
\begin{align*}
sid_1 \xrightarrow{\text{since}} 2020 : sid_4
\end{align*}
to be ``invisible'' in the context of a plain RDF model; alternatively, we could have represented this assertion using a set of triples that leverage the RDF reification vocabulary. The goal of this paper is not to provide a complete formalization of the 1G model and its interpretation, but rather leverage it as a framework to discuss the key questions that need to be answered in order to achieve graph interoperability, and would lead to a “natural” semantics for the query languages over a unified graph data model.\footnote{In Section~\ref{namedgraphs} we show how the 1G model is used to represent RDF's {\em named graphs}.}
The discussion in this paper is limited to LPGs used by Gremlin and openCypher. Other graph query languages and implementations may come with additional challenges to consider. Our evaluation of RDF (wrt. our 1G model) includes features of RDF-star.

\section{Interoperability challenges and some proposed solutions}

We discuss challenges to interoperability broadly, ranging from semantics to potential implementation issues.\footnote{Note that the issue of {\em schema languages} is out of scope for this paper.}

\subsection{Challenge \#1: Edge properties, multiple edge instances, and reification}
\label{reification}

One of the most fundamental perceived (structural) difference between RDF and LPGs is that RDF does not offer built-in edge property support (except as “statements about statements” via reification, which is generally understood to be cumbersome and inefficient). We sketched this aspect in Section~\ref{1G} already when talking about possible interpretations for edge properties in RDF.  A related issue is that LPGs support more than one instance of an edge. The RDF-star effort is adding edge properties to RDF, but – as the draft specification currently stands – not in a way that is completely compatible with LPGs. In RDF-star as currently formulated, a triple $<s, p, o>$ is understood to be unique, in the sense that there cannot exist an identical instance of that triple with its own identity. As such, all statements involving said $<s, p, o>$ are understood to reference the {\em same} triple (in contrast, the original RDF reification mechanism \cite[section 4.3]{Schreiber+Raimond-RDFPrimer2014} is more expressive: each statement model has its own identity). In LPGs, each edge is considered a unique object with its own identity, and it is perfectly possible to have two edges with identical endpoints and label. The following example illustrates the case of multiple edge identities using the 1G model: 
\begin{align*}
&\texttt{Alice} \xrightarrow{\text{knows}} \texttt{Bob} : sid_1\\
&\texttt{Alice} \xrightarrow{\text{knows}} \texttt{Bob} : sid_2
\end{align*}
Further assume there are four assertions over these two ground edges:
\begin{align*}
&sid_1 \xrightarrow{\text{statedBy}} \texttt{NYTimes} : sid_3\\
&sid_1 \xrightarrow{\text{since}} 2020 : sid_4\\
&sid_2 \xrightarrow{\text{statedBy}} \texttt{TheGuardian} : sid_5\\
&sid_2 \xrightarrow{\text{since}} 2021 : sid_6
\end{align*}
Informally speaking, the four assertions over the ground facts state that Alice knows Bob (a) since 2020 as per the NY Times, and (b) since 2021 as per The Guardian. This example has a natural representation in LPGs (namely, two distinguished edges, each carrying the respective two edge properties), whereas there is no built-in mechanism in RDF-star as defined at the time of writing. This means that with respect to edge properties (and some other issues notwithstanding) we can take an RDF(-star) graph and map it onto an LPG, but there are cases where the converse is not true, at least not without falling back on more complex modeling approaches (such as introducing intermediate nodes or using named graphs to attach custom statement identifiers). 

Prohibiting multiple edges would seriously hamper how LPGs are used -- there is a clear use case behind the distinction of the two edges in the example above -- so it makes sense to capture this expressiveness in 1G. The question then becomes what an RDF-star view (where the difference cannot be expressed) over such data would look like. One option is to collapse multiple edges (in our example, $sid_1$ and $sid_2$) into a single edge, effectively exposing a single edge with four edge properties when looking at 1G from an RDF-star perspective. This may still allow for some useful SPARQL-star queries (e.g., a query such as ``give me all ground facts stated by \texttt{:NYTimes}'') but in the general case information would be lost. Conceptually, this could be seen as a ``dimensional reduction'' when mapping 1G into RDF-star, which is accounting for the fact that RDF-star is just not expressive enough to capture the distinction between these two edges.

There are a couple of other, similar questions related to reification, apart from the edge identity challenge discussed above. Those include aspects such as multi-level reification (which can be expressed with RDF-star but not with LPGs), questions whether the reification of a 1G statement via its $sid$ necessarily implies the presence of the underlying statement as a ground statement or not (in our 1G model it does, and in fact the model is not capable of referring to non-asserted statements). While those may sound like technical details, we believe that the challenge of overcoming these ``reification gaps'' is at the very heart of graph interoperability. 

From our experience, the convenience of having edge properties as a built-in construct is one of the core strengths of LPGs over RDF. Given the existing gaps in expressiveness and the early stage of RDF-star proposals, it might also be worth re-considering RDF-star and better aligning its expressiveness to capture the distinction between multiple edges.

\subsection{Challenge \#2: Triples vs. graph abstraction}

At the very basic level, an RDF graph is defined as a ``set of triples'', whereas LPGs are defined as (optionally labeled) nodes with properties that can be connected via labeled edges. The LPG notion of a {\em vertex label} is very similar to that of \texttt{rdf:type}.  That is, LPG labels classify a vertex.  While the mapping of triples to and from our abstract 1G model is straightforward (in the absence of reification, there is one 1G statement for every triple, with a unique {\em sid} assigned),
the mapping from LPGs into the 1G model is more challenging. For instance, certain flavors of LPGs allow ``stand-alone'' nodes without labels and properties, i.e., a node itself that neither carries any information nor is connected to any other node.\footnote{Note that RDF, effectively, {\em does not have vertices}, in the sense that they could exist ``stand-alone''. There is an infinite space of identifiers, any of which could be used as vertices. This is not merely a philosophical notion, as we now see.} In fact, such nodes can be extracted via query languages (in Gremlin via \texttt{g.V()} or in openCypher via \texttt{MATCH (n)}), but there is no natural representation for these stand-alone nodes in our 1G model. In Gremlin and openCypher, vertices always have at least a default label, however.\footnote{This is not unlike every node in an RDF graph being an instance of \texttt{rdfs:Resource}.} 

Similar questions arise when querying data originating from RDF graphs via LPG query languages. LPG languages typically follow the abstraction layer of vertices and edges. Gremlin, for instance, allows to enumerate all vertices using the query \texttt{g.V()} whereas edges can be extracted via \texttt{g.E()}. In the LPG model, the sets of vertices and edges are {\em disjoint}.

The key challenge in defining the semantics of LPG query languages becomes defining the concepts of vertices and edges over our 1G model -- which would manifest itself in a formal mapping from 1G to LPGs. Concretely, the question becomes how we define the concepts of vertices, edges, vertex properties, and edge properties in a 1G-to-LPG mapping. The basic idea here would be to map identifiers from \texttt{src} and \texttt{value} position of 1G statements to vertices, and use the {\em sids} as edge or property identifiers (depending on whether a statement carries an identifier (such as \texttt{Alice} or \texttt{Bob}) or a literal (such as a string \texttt{"Alice"} or \texttt{"Bob"}) in its \texttt{value} position, respectively).

\subsection{Challenge \#3: Datatype alignment}

1G requires a unified type system over RDF and LPGs. As a W3C standard it builds upon the XML Schema definition (XSD) \cite{XSD-2004} and utilizes primitive XSD datatypes such as strings, numbers, and dates. Since RDF is defined along an Open World paradigm, its datatypes tend to be more extensible and flexible than LPG types. For instance, there is no validation in place that restricts users from adding ill-typed values such as \texttt{"this is not an integer"$\wedge\wedge$xsd:integer}, and users can also provide language tags for simple string values. Composite types such as lists, bags, and sequences are not available as ``primary'' literal types but need to be modeled explicitly using RDF containers.\footnote{\url{https://www.w3.org/TR/rdf-schema/\#ch\_container}} RDF supports bags, but not sets, because the graph structure offers no easy means to enforce set semantics while also permitting the merge of RDF data sets. 

For LPGs, on the other hand, there exists (to the best of our knowledge) no formal definition of the type system. In contrast to RDF, LPGs support different sets of composite datatypes as built-in types. This means that, while RDF uses the graph structure to model composite types, in LPGs the attribute value itself is an instance of a composite type (a list, map, etc.). This reflects the general notion of semi-structured JSON documents as attribute values. Semantics of these datatypes, however, are opaque and are typically ``delegated'' to the underlying implementation language, making it potentially hard to unify graph representation. Generally, we have a menagerie of datatypes to try to reconcile, and needless ``baggage'' because of the reliance on implementation languages. The challenge for a unifying graph data model is to define a {\em meta type system} that captures and aligns these different types, giving them a concrete semantics.

One possible approach to align the type systems between RDF and LPGs more closely -- with a focus on composite types -- would be to leverage RDF’s user-defined literal mechanism, which is an extension mechanism that can be used to represent arbitrary complex types. As an alternative to the structural approach usually taken by RDF, a list could be represented in RDF as a literal value whose lexical form is, say, \texttt{"[1, 2, 3]"$\wedge\wedge$:OneGraphList}, the serialized form of a list, and where \texttt{:OneGraphList} is a reserved datatype IRI defining the literal value to be a list with well-defined syntax that can be used interchangeably in RDF and LPGs.\footnote{The syntax makes no assertion about how the data are represented internally.} This approach renders these datatypes opaque with respect to existing RDF semantics, but for processing (say, in SPARQL) we could define some type of folding/unfolding operations to expose the internal structure, or even specific user-defined functions such as element projections that understand and interpret such typed literals. In some ways these operations could be considered an extension of the ``literal value'' notion that is part of the RDF specification \cite[section 3.3]{Klyne+Carroll-RDFSpec2004}. 

\subsection{Challenge \#4: Graph partitioning}
\label{namedgraphs}

SPARQL defines the notion of {\em named graphs}, which are often used to support subgraph management use cases. Named graphs are usually thought of as an extension of the triple model to a quad model with the addition of a (sub)graph identifier. Some users have chosen to treat named graphs as containers (sometimes containers of a single triple) to make ``statements about statements'' (or sets of statements) in lieu of using the reification mechanism (this, in the absence of the proposed RDF-star scenario). This is outside the defined formal semantics of RDF, since named graphs do not have any semantic theory in the RDF model.  

Rather than having an RDF quad like $<s, p, o, g>$, in~1G~we introduce a ``membership relation'' (in this paper we use \texttt{inGraph} as a reserved label):
\begin{flalign*}
&s  \xrightarrow{\text{p}} o : sid_1\\
&sid_1  \xrightarrow{\text{inGraph}} g : sid_2
\end{flalign*}
where $sid_1$ and $sid_2$ are statement identifiers. The motivation behind this approach is to restore symmetry to the data model instead of privileging named graphs as somehow special. Named graphs are just an application of the statement identifier, much like {\em aspects} in \cite{Lassila+etal-Aspects2010}.  Note, however, that the 1G model does not dictate how a database physically organizes the graph. There is, of course, a storage size consideration when using additional triples/quads.

From the standpoint of SPARQL semantics, the membership triples are not visible, but instead could be considered an implementation artifact.  However, with this proposal they would be part of the data in the data set.  SPARQL default graph and named graph semantics are recovered by performing the appropriate operations over the logical model. Again, we do not imply that the physical schema needs to have this same data organization.

Since in our proposed solution we treat named graph membership effectively as an ``edge or meta property'', this approach can be extended to LPGs. In an LPG implementation where properties (including vertex properties) are represented as triples/quads, this approach permits vertex properties to be assigned graph membership. It remains a debatable issue whether the notion of subgraphs in an LPG would extend to vertices themselves and, if not, how the LPG view would be recovered from the 1G model. Obviously in RDF this is not an issue since RDF vertices are merely names with no internal structure.

\subsection{Challenge \#5: Graph merging and external identifiers}
\label{merging}

This is one of the strongest benefits of RDF, and conversely one of the weakest aspects of LPGs: RDF has a specification for graph merging \cite[section 4.1]{Hayes-RDFMT2004}. Whenever multiple data sources are used, particularly from multiple organizations, such as in the Linked Open Data cloud, graph merging is a key functionality. Strong external identifiers (IRIs) are an essential part of this mechanism, but while it is perfectly possible to use IRIs as identifiers in LPGs, there is more to graph merging; edge properties, and particularly multiple edge instances (see above) complicate any merging semantics. Specifically, under which conditions could two ``similar'' edges in two graphs that are to be merged be considered the ``same'' edge?

Allowing both RDF and LPG data to be represented in a single model requires the co-existence of global identifiers (i.e., IRIs coming from RDF data) and local identifiers (i.e., node and edge identifiers in the form of strings, coming from LPG data). Such a distinction would make it possible to load data via both RDF and LPG data formats into the same logical graph; this would give us mere co-existence, without any (initial) overlap in vertex identifiers, labels, etc. User-defined default namespaces could then be used to expose local identifiers as if they were IRIs, whenever they are queried via SPARQL; conversely, IRIs originating from RDF data sets could be shortened according to existing namespace prefixes, to make querying and syntactic result representation for LPG query languages more user-friendly.

The typical use case, however, goes beyond just the mere co-existence of local and global identifiers: a common use case would be to unify elements (such as nodes and edges) originating from RDF and LPGs. Assume, for instance, a user who maintains internal data about countries, available as LPG data in which countries are identified via a simple country code string, and that the user wants to augment this data with information coming from an RDF dump of the Geonames dataset.\footnote{\url{https://www.geonames.org/}} When initially loading the data sets, the LPG country identifier strings and the IRIs for the respective countries coming from Geonames would be disjoint; in order to allow users to align such identifiers, we would need to provide user-configurable rules that guide the merging process (in this case, collating the country IRIs from the Geonames RDF data with the country code strings from the LPG dataset). This mechanism could also be used to customize RDF graph merging when blank nodes are present (e.g., when two graphs derived from the same source are merged, one may want to treat blank node identifiers as indeed identifying the same blank nodes – this is something that is allowed under RDF semantics). Identifiers could also be generated using a templating mechanism such as the one employed in R2RML \cite[section 7.3]{Das+etal-R2RML2012}.

\subsection{Challenge \#6: Lack of formal foundation}

Unlike SPARQL or SQL, LPG query languages -- by and large -- lack strict formal semantics (in the form of, say, a query algebra). This makes it hard to assess semantic compatibility. Similarly, unlike RDF, formal semantics do not exist for LPGs (or only exist {\em post hoc}, as in \cite{shinavier2019,marton2017}). For LPG query languages, semantics is typically defined informally either via documentation and examples, or via an implementation. Our Gremlin implementation, for instance, is largely based on our interpretation of the informal Tinkerpop specification\footnote{\url{https://tinkerpop.apache.org/gremlin.html}} -- which comes with details that are not unambiguously defined -- and its reference implementation. To provide a simple example, consider the question of ordering guarantees: if a Gremlin query contains an \texttt{order()} step in a non-terminal position, it is ambiguous whether (and how) subsequent steps should maintain that order. For example, would a Gremlin query such as \texttt{g.V(1,2,3,4).order().by('age').out()} guarantee that \texttt{out()} nodes have to be reported for nodes with ids 1, 2, 3, and 4 in the order implied by their age property?
Similar questions around semantics arise in the context of dynamic typing, execution order guarantees, and runtime exceptions thrown through the query language. And of course, such questions arise in the context of other query languages like openCypher.  While openCypher does provide an official spec, certain details are not entirely worked out. Of course, SPARQL is not entirely without fault either, as there is some ambiguity in its specification as well -- see, for example, \cite{hernandez2016} and also \cite{patel2016}.

\subsection{Challenge \#7: Update query semantics}

In order to be able to subsume both the RDF(-star) and the LPG models, a unifying graph model needs to be as expressive as the ``most expressive'' model, in each of the considered dimensions. As we have illustrated in previous examples, certain extensions that are defined for the more expressive model may not have a natural representation in the less expressive model, thus introducing ``dimensions'' that are invisible when looking at the data from the less expressive model’s perspective. As a concrete example, we discussed in Section~\ref{reification} that multiple edge properties modeled in our 1G model, while visible in an LPG interpretation, may collapse into a single edge property in an RDF-star interpretation.

While read query semantics can be unambiguously defined by mapping the 1G model to a lower-dimensional level, the situation becomes more complex for queries that manipulate the data. Consider, for instance, the example database sketched in challenge \#1, and assume we write a SPARQL query (non- SPARQL-star) that requests to drop the triple \texttt{:Bob :knows :Alice}. From a perspective of SPARQL (e.g., when querying), this triple would always show up once, but the question is whether dropping it should -- behind the scenes -- delete both ground statements? And if so, should deletion include the deletion of the attached edge properties (which exposes a risk for accidental over-deletion)? Would adding an edge for every subject with IRI \texttt{:Bob} to a new person introduce one or two edges?
As another scenario, when we use SPARQL-star and add a new edge property to the toy 1G model introduced in Section~\ref{reification}, which of the two edges would we attach it to?

The root cause for the ambiguity of all these scenarios is that we request updates over a simplified, dimensionally reduced view of the data. While the discussion of all these questions goes beyond the scope of this paper, the semantics of such operations needs to be carefully designed in order to avoid unwanted side effects and obtain a ``natural'' behavior in scenarios where data is queried and manipulated via different query languages. 

\section{Conclusions and the way forward}

A recent survey \cite{EKGF-survey2021} of organizations working on (or considering to adopt) knowledge graphs found that {\em interoperability and standards} are the highest priority among survey respondents. {\em Data integration} was seen as the dominant use case. These findings could be interpreted to suggest a need for RDF/LPG compatibility and unification. While it may seem that making RDF and LPGs fully compatible is not possible (as per the official RDF specifications and the emerging RDF-star work), we believe there is a way forward. Minimally, we must address the challenges of edge identity (multiple similar edges), graph merging, and well-defined semantics for updates across languages. One way forward would be to define some kind of ``compatibility subset'' to cover enough ground so that most RDF and LPG applications would work with no or minimal modifications. Lack of interoperability slows the overall adoption of graph technologies, and thus should be a high-priority item to be addressed by the broader graph community.

In this paper we have presented the 1G model as a source of ideas about where to go next. The open questions we have posed imply lots of interesting research, the outcomes of which will have significant practical relevance. We now seek support from the broader community to look into these topics.

\bibliographystyle{splncs04}
\bibliography{SemanticWeb}

\begin{thebibliography}{10}
\providecommand{\url}[1]{\texttt{#1}}
\providecommand{\urlprefix}{URL }
\providecommand{\doi}[1]{https://doi.org/#1}

\bibitem{EKGF-survey2021}
Atkin, M., Deely, T., Scharffe, F.: {Knowledge Graph Benchmarking Report 2021
  (version 2.0)}. Zenodo, http://doi.org/10.5281/zenodo.4950097 (June 2021)

\bibitem{XSD-2004}
Biron, P.V., Malhotra, A.: {XML Schema Part 2: Datatypes (Second Edition)}.
  {W3C Recommendation}, {World Wide Web Consortium} (Oct 2004),
  \url{http://www.w3.org/TR/2004/REC-xmlschema-2-20041028/}

\bibitem{Das+etal-R2RML2012}
Das, S., Sundara, S., Cyganiak, R.: {R2RML: RDB to RDF Mapping Language}. {W3C
  Recommendation}, World Wide Web Consortium, Cambridge, MA (Sep 2012)

\bibitem{Harris+Seaborne-SPARQLSpec2013}
Harris, S., Seaborne, A.: {SPARQL 1.1 Query Language}. {W3C Recommendation},
  World Wide Web Consortium, Cambridge, MA (2013)

\bibitem{hartig2021}
Hartig, O., Champin, P.A., Kellogg, G., Seaborne, A., Arndt, D., Broekstra, J.,
  DuCharme, B., Lassila, O., Patel-Schneider, P.F., Prud'hommeaux, E.,
  Thibodeau, Jr., T., Bryan, T.: {RDF-star and SPARQL-star}. {W3C Community
  Group Working Draft}, World Wide Web Consortium (June 2021)

\bibitem{Hayes-RDFMT2004}
Hayes, P.: {RDF Semantics}. {W3C Recommendation}, World Wide Web Consortium
  (Feb 2004), \url{http://www.w3.org/TR/rdf-mt/}

\bibitem{hernandez2016}
Hern\'andez, D., Gutierrez, C., Angles, R.: {Correlation and Substitution in
  SPARQL}. arXiv preprint arXiv:1606.01441  (2016)

\bibitem{Klyne+Carroll-RDFSpec2004}
Klyne, G., Carroll, J.J.: {Resource Description Framework (RDF): Concepts and
  Abstract Syntax}. {W3C Recommendation}, World Wide Web Consortium, Cambridge,
  MA (Feb 2004), \url{http://www.w3.org/TR/rdf-concepts/}

\bibitem{Lassila+etal-Aspects2010}
Lassila, O., Mannermaa, M., Oliver, I., Sabbouh, M.: {Aspect-Oriented Data}.
  {Accepted submission to the W3C Workshop on RDF Next Steps}, {World Wide Web
  Consortium} (June 2010)

\bibitem{marton2017}
Marton, J., Sz{\'a}rnyas, G., Varr{\'o}, D.: {Formalising openCypher graph
  queries in relational algebra}. In: European Conference on Advances in
  Databases and Information Systems. pp. 182--196. Springer (2017)

\bibitem{patel2016}
Patel-Schneider, P.F., Martin, D.: {EXISTStential Aspects of SPARQL}. In:
  International Semantic Web Conference (Posters \& Demos). vol. CEUR-WS \#1690
  (2016)

\bibitem{Prudhommeaux+Buil-Aranda-SPARQLSpec2013}
Prud'hommeaux, E., Buil-Aranda, C.: {SPARQL 1.1 Federated Query}. {W3C
  Recommendation}, World Wide Web Consortium, Cambridge, MA (2013)

\bibitem{robinson2013}
Robinson, I., Webber, J., Eifrem, E.: Graph databases: new opportunities for
  connected data. O'Reilly Media, Inc. (2013)

\bibitem{rodriguez2015}
Rodriguez, M.A.: {The Gremlin Graph Traversal Machine and Language (invited
  talk)}. In: Proceedings of the 15th Symposium on Database Programming
  Languages (DBPL 2015). pp. 1--10 (2015)

\bibitem{rodriguez2010}
Rodriguez, M.A., Neubauer, P.: Constructions from dots and lines. Bulletin of
  the American Society for Information Science and Technology  \textbf{36}(6),
  35--41 (2010)

\bibitem{Schreiber+Raimond-RDFPrimer2014}
Schreiber, G., Raimond, Y.: {RDF 1.1 Primer}. {W3C Recommendation}, World Wide
  Web Consortium, Cambridge, MA (2014)

\bibitem{shinavier2019}
Shinavier, J., Wisnesky, R.: Algebraic property graphs. arXiv preprint
  arXiv:1909.04881  (2019)

\end{thebibliography}

\end{document}